%% file: 21ASIL.tex
\documentclass[conference,10pt,twocolumn,letter]{IEEEtran}
\IEEEoverridecommandlockouts
\usepackage{cite}
\usepackage[T1]{fontenc}
\usepackage{graphicx}
\usepackage{amssymb}
\usepackage{amsmath}
\usepackage{amsthm}

\usepackage{microtype}
\usepackage{balance}
\usepackage{subfigure}
\input{vmr-symbols-vecbold}
\input{standard-macros}

\input{defs}

\newcommand*{\fancyrefremarklabelprefix}{remark}
\frefformat{vario}{\fancyrefremarklabelprefix}{Remark~#1}

\IEEEoverridecommandlockouts % enables \thanks in conference mode
\allowdisplaybreaks %%%% THIS THING ALLOWS EQUATIONS TO BREAK THE PAGE (saves tons of space)
\usepackage{color}
\newcommand{\revision}[1]{#1}

\usepackage{bbm}

\def\bSigma{\mathbf{\Sigma}}

\newcommand{\PARpq}[2]{\textit{PAR}_{#2}^{#1}}

\makeatletter

\newcommand{\leqnomode}{\tagsleft@true}
\newcommand{\reqnomode}{\tagsleft@false}

\makeatother

\title{$\ell^p\!-\!\ell^q$-Norm Minimization for Joint Precoding \\ and Peak-to-Average-Power Ratio Reduction}
\author{\IEEEauthorblockN{Sueda Taner$^\text{1}$ and Christoph Studer$^\text{2}$} \\[0.05cm]
\thanks{The work of ST and CS was supported by ComSenTer, one of six centers in JUMP, an SRC program sponsored by DARPA. The work of CS was also supported in part by an ETH Research Grant and by the US NSF under grants CNS-1717559 and ECCS-1824379.}
\thanks{The authors would like to thank C.~Jeon, O. Casta\~neda, C.~Dick, T.~Goldstein, and E.~Larsson for discussions on $\ell^\infty$-norm-based PAR reduction.}
\IEEEauthorblockA{$^\text{1}$School of Electrical and Computer Engineering, Cornell University, Ithaca, NY;  e-mail: st939@cornell.edu\\[0.02cm]
$^\text{2}$Department of Information Technology and Electrical Engineering, ETH Z\"urich, Switzerland; e-mail: studer@ethz.ch} 
}

\begin{document}

\maketitle

\input{0-abstract}

\input{1-introduction}
\input{2-prereqs}
\input{3ab-problem}

\input{3cd-solution}

\input{4-ofdm}

\input{5-results}

\input{6-conclusion}

 \input{7-appendix}

\bibliographystyle{IEEEtran}

\linespread{0.983} 
\bibliography{bib/IEEEabrv,bib/confs-jrnls,bib/publishers,bib/studer,bib/vipbib,21ASIL_bib} %,bib/VIP

\balance
	
\end{document}

%% file: vmr-symbols-vecbold.tex
\usepackage{amssymb}
\usepackage{amsfonts}
\usepackage{mathrsfs}
\usepackage{xspace}
\usepackage{bm}
\usepackage{upgreek}

\newcommand{\safemath}[2]{\newcommand{#1}{\ensuremath{#2}\xspace}}

\safemath{\bma}{\mathbf{a}}
\safemath{\bmb}{\mathbf{b}}
\safemath{\bmc}{\mathbf{c}}
\safemath{\bmd}{\mathbf{d}}
\safemath{\bme}{\mathbf{e}}
\safemath{\bmf}{\mathbf{f}}
\safemath{\bmg}{\mathbf{g}}
\safemath{\bmh}{\mathbf{h}}
\safemath{\bmi}{\mathbf{i}}
\safemath{\bmj}{\mathbf{j}}
\safemath{\bmk}{\mathbf{k}}
\safemath{\bml}{\mathbf{l}}
\safemath{\bmm}{\mathbf{m}}
\safemath{\bmn}{\mathbf{n}}
\safemath{\bmo}{\mathbf{o}}
\safemath{\bmp}{\mathbf{p}}
\safemath{\bmq}{\mathbf{q}}
\safemath{\bmr}{\mathbf{r}}
\safemath{\bms}{\mathbf{s}}
\safemath{\bmt}{\mathbf{t}}
\safemath{\bmu}{\mathbf{u}}
\safemath{\bmv}{\mathbf{v}}
\safemath{\bmw}{\mathbf{w}}
\safemath{\bmx}{\mathbf{x}}
\safemath{\bmy}{\mathbf{y}}
\safemath{\bmz}{\mathbf{z}}
\safemath{\bmzero}{\mathbf{0}}
\safemath{\bmone}{\mathbf{1}}

\bmdefine{\biad}{a}
\bmdefine{\bibd}{b}
\bmdefine{\bicd}{c}
\bmdefine{\bidd}{d}
\bmdefine{\bied}{e}
\bmdefine{\bifd}{f}
\bmdefine{\bigd}{g}
\bmdefine{\bihd}{h}
\bmdefine{\biid}{i}
\bmdefine{\bijd}{j}
\bmdefine{\bikd}{k}
\bmdefine{\bild}{l}
\bmdefine{\bimd}{m}
\bmdefine{\bind}{n}
\bmdefine{\biod}{o}
\bmdefine{\bipd}{p}
\bmdefine{\biqd}{q}
\bmdefine{\bird}{r}
\bmdefine{\bisd}{s}
\bmdefine{\bitd}{t}
\bmdefine{\biud}{u}
\bmdefine{\bivd}{v}
\bmdefine{\biwd}{w}
\bmdefine{\bixd}{x}
\bmdefine{\biyd}{y}
\bmdefine{\bizd}{z}

\bmdefine{\bixid}{\xi}
\bmdefine{\bilambdad}{\lambda}
\bmdefine{\bimud}{\mu}
\bmdefine{\bithetad}{\theta}
\bmdefine{\biphid}{\phi}
\bmdefine{\bideltad}{\delta}

\safemath{\bmia}{\biad}
\safemath{\bmib}{\bibd}
\safemath{\bmic}{\bicd}
\safemath{\bmid}{\bidd}
\safemath{\bmie}{\bied}
\safemath{\bmif}{\bifd}
\safemath{\bmig}{\bigd}
\safemath{\bmih}{\bihd}
\safemath{\bmii}{\biid}
\safemath{\bmij}{\bijd}
\safemath{\bmik}{\bikd}
\safemath{\bmil}{\bild}
\safemath{\bmim}{\bimd}
\safemath{\bmin}{\bind}
\safemath{\bmio}{\biod}
\safemath{\bmip}{\bipd}
\safemath{\bmiq}{\biqd}
\safemath{\bmir}{\bird}
\safemath{\bmis}{\bisd}
\safemath{\bmit}{\bitd}
\safemath{\bmiu}{\biud}
\safemath{\bmiv}{\bivd}
\safemath{\bmiw}{\biwd}
\safemath{\bmix}{\bixd}
\safemath{\bmiy}{\biyd}
\safemath{\bmiz}{\bizd}

\safemath{\bmxi}{\bixid}
\safemath{\bmlambda}{\bilambdad}
\safemath{\bmmu}{\bimud}
\safemath{\bmtheta}{\bithetad}
\safemath{\bmphi}{\biphid}
\safemath{\bmdelta}{\bideltad}

\safemath{\bA}{\mathbf{A}}
\safemath{\bB}{\mathbf{B}}
\safemath{\bC}{\mathbf{C}}
\safemath{\bD}{\mathbf{D}}
\safemath{\bE}{\mathbf{E}}
\safemath{\bF}{\mathbf{F}}
\safemath{\bG}{\mathbf{G}}
\safemath{\bH}{\mathbf{H}}
\safemath{\bI}{\mathbf{I}}
\safemath{\bJ}{\mathbf{J}}
\safemath{\bK}{\mathbf{K}}
\safemath{\bL}{\mathbf{L}}
\safemath{\bM}{\mathbf{M}}
\safemath{\bN}{\mathbf{N}}
\safemath{\bO}{\mathbf{O}}
\safemath{\bP}{\mathbf{P}}
\safemath{\bQ}{\mathbf{Q}}
\safemath{\bR}{\mathbf{R}}
\safemath{\bS}{\mathbf{S}}
\safemath{\bT}{\mathbf{T}}
\safemath{\bU}{\mathbf{U}}
\safemath{\bV}{\mathbf{V}}
\safemath{\bW}{\mathbf{W}}
\safemath{\bX}{\mathbf{X}}
\safemath{\bY}{\mathbf{Y}}
\safemath{\bZ}{\mathbf{Z}}

\safemath{\bZero}{\mathbf{0}}
\safemath{\bOne}{\mathbf{1}}
\safemath{\bDelta}{\mathbf{\Delta}}
\safemath{\bLambda}{\mathbf{\UpLambda}}
\safemath{\bPhi}{\mathbf{\Upphi}}
\safemath{\bSigma}{\mathbf{\Upsigma}}
\safemath{\bOmega}{\mathbf{\Upomega}}
\safemath{\bTheta}{\mathbf{\Uptheta}}

\bmdefine{\biAd}{A}
\bmdefine{\biBd}{B}
\bmdefine{\biCd}{C}
\bmdefine{\biDd}{D}
\bmdefine{\biEd}{E}
\bmdefine{\biFd}{F}
\bmdefine{\biGd}{G}
\bmdefine{\biHd}{H}
\bmdefine{\biId}{I}
\bmdefine{\biJd}{J}
\bmdefine{\biKd}{K}
\bmdefine{\biLd}{L}
\bmdefine{\biMd}{M}
\bmdefine{\biOd}{N}
\bmdefine{\biPd}{O}
\bmdefine{\biQd}{P}
\bmdefine{\biRd}{R}
\bmdefine{\biSd}{S}
\bmdefine{\biTd}{T}
\bmdefine{\biUd}{U}
\bmdefine{\biVd}{V}
\bmdefine{\biWd}{W}
\bmdefine{\biXd}{X}
\bmdefine{\biYd}{Y}
\bmdefine{\biZd}{Z}

\bmdefine{\biDelta}{\Delta}
\bmdefine{\biLambda}{\Lambda}
\bmdefine{\biPhi}{\Phi}
\bmdefine{\biSigma}{\Sigma}
\bmdefine{\biOmega}{\Omega}
\bmdefine{\biTheta}{\Theta}

\safemath{\bimA}{\biAd}
\safemath{\bimB}{\biBd}
\safemath{\bimC}{\biCd}
\safemath{\bimD}{\biDd}
\safemath{\bimE}{\biEd}
\safemath{\bimF}{\biFd}
\safemath{\bimG}{\biGd}
\safemath{\bimH}{\biHd}
\safemath{\bimI}{\biId}
\safemath{\bimJ}{\biJd}
\safemath{\bimK}{\biKd}
\safemath{\bimL}{\biLd}
\safemath{\bimM}{\biMd}
\safemath{\bimN}{\biNd}
\safemath{\bimO}{\biOd}
\safemath{\bimP}{\biPd}
\safemath{\bimQ}{\biQd}
\safemath{\bimR}{\biRd}
\safemath{\bimS}{\biSd}
\safemath{\bimT}{\biTd}
\safemath{\bimU}{\biUd}
\safemath{\bimV}{\biVd}
\safemath{\bimW}{\biWd}
\safemath{\bimX}{\biXd}
\safemath{\bimY}{\biYd}
\safemath{\bimZ}{\biZd}

\safemath{\bimDelta}{\biDelta}
\safemath{\bimLambda}{\biLambda}
\safemath{\bimPhi}{\biPhi}
\safemath{\bimSigma}{\biSigma}
\safemath{\bimOmega}{\biOmega}
\safemath{\bimTheta}{\biTheta}

\safemath{\setA}{\mathcal{A}}
\safemath{\setB}{\mathcal{B}}
\safemath{\setC}{\mathcal{C}}
\safemath{\setD}{\mathcal{D}}
\safemath{\setE}{\mathcal{E}}
\safemath{\setF}{\mathcal{F}}
\safemath{\setG}{\mathcal{G}}
\safemath{\setH}{\mathcal{H}}
\safemath{\setI}{\mathcal{I}}
\safemath{\setJ}{\mathcal{J}}
\safemath{\setK}{\mathcal{K}}
\safemath{\setL}{\mathcal{L}}
\safemath{\setM}{\mathcal{M}}
\safemath{\setN}{\mathcal{N}}
\safemath{\setO}{\mathcal{O}}
\safemath{\setP}{\mathcal{P}}
\safemath{\setQ}{\mathcal{Q}}
\safemath{\setR}{\mathcal{R}}
\safemath{\setS}{\mathcal{S}}
\safemath{\setT}{\mathcal{T}}
\safemath{\setU}{\mathcal{U}}
\safemath{\setV}{\mathcal{V}}
\safemath{\setW}{\mathcal{W}}
\safemath{\setX}{\mathcal{X}}
\safemath{\setY}{\mathcal{Y}}
\safemath{\setZ}{\mathcal{Z}}
\safemath{\emptySet}{\varnothing}

\safemath{\colA}{\mathscr{A}}
\safemath{\colB}{\mathscr{B}}
\safemath{\colC}{\mathscr{C}}
\safemath{\colD}{\mathscr{D}}
\safemath{\colE}{\mathscr{E}}
\safemath{\colF}{\mathscr{F}}
\safemath{\colG}{\mathscr{G}}
\safemath{\colH}{\mathscr{H}}
\safemath{\colI}{\mathscr{I}}
\safemath{\colJ}{\mathscr{J}}
\safemath{\colK}{\mathscr{K}}
\safemath{\colL}{\mathscr{L}}
\safemath{\colM}{\mathscr{M}}
\safemath{\colN}{\mathscr{N}}
\safemath{\colO}{\mathscr{O}}
\safemath{\colP}{\mathscr{P}}
\safemath{\colQ}{\mathscr{Q}}
\safemath{\colR}{\mathscr{R}}
\safemath{\colS}{\mathscr{S}}
\safemath{\colT}{\mathscr{T}}
\safemath{\colU}{\mathscr{U}}
\safemath{\colV}{\mathscr{V}}
\safemath{\colW}{\mathscr{W}}
\safemath{\colX}{\mathscr{X}}
\safemath{\colY}{\mathscr{Y}}
\safemath{\colZ}{\mathscr{Z}}

\safemath{\opA}{\mathbb{A}}
\safemath{\opB}{\mathbb{B}}
\safemath{\opC}{\mathbb{C}}
\safemath{\opD}{\mathbb{D}}
\safemath{\opE}{\mathbb{E}}
\safemath{\opF}{\mathbb{F}}
\safemath{\opG}{\mathbb{G}}
\safemath{\opH}{\mathbb{H}}
\safemath{\opI}{\mathbb{I}}
\safemath{\opJ}{\mathbb{J}}
\safemath{\opK}{\mathbb{K}}
\safemath{\opL}{\mathbb{L}}
\safemath{\opM}{\mathbb{M}}
\safemath{\opN}{\mathbb{N}}
\safemath{\opO}{\mathbb{O}}
\safemath{\opP}{\mathbb{P}}
\safemath{\opQ}{\mathbb{Q}}
\safemath{\opR}{\mathbb{R}}
\safemath{\opS}{\mathbb{S}}
\safemath{\opT}{\mathbb{T}}
\safemath{\opU}{\mathbb{U}}
\safemath{\opV}{\mathbb{V}}
\safemath{\opW}{\mathbb{W}}
\safemath{\opX}{\mathbb{X}}
\safemath{\opY}{\mathbb{Y}}
\safemath{\opZ}{\mathbb{Z}}
\safemath{\opZero}{\mathbb{O}}
\safemath{\identityop}{\opI}

\safemath{\veca}{\bma}
\safemath{\vecb}{\bmb}
\safemath{\vecc}{\bmc}
\safemath{\vecd}{\bmd}
\safemath{\vece}{\bme}
\safemath{\vecf}{\bmf}
\safemath{\vecg}{\bmg}
\safemath{\vech}{\bmh}
\safemath{\veci}{\bmi}
\safemath{\vecj}{\bmj}
\safemath{\veck}{\bmk}
\safemath{\vecl}{\bml}
\safemath{\vecm}{\bmm}
\safemath{\vecn}{\bmn}
\safemath{\veco}{\bmo}
\safemath{\vecp}{\bmp}
\safemath{\vecq}{\bmq}
\safemath{\vecr}{\bmr}
\safemath{\vecs}{\bms}
\safemath{\vect}{\bmt}
\safemath{\vecu}{\bmu}
\safemath{\vecv}{\bmv}
\safemath{\vecw}{\bmw}
\safemath{\vecx}{\bmx}
\safemath{\vecy}{\bmy}
\safemath{\vecz}{\bmz}

\safemath{\veczero}{\bmzero}
\safemath{\vecone}{\bmone}
\safemath{\vecxi}{\bmxi}
\safemath{\veclambda}{\bmlambda}
\safemath{\vecmu}{\bmmu}
\safemath{\vectheta}{\bmtheta}
\safemath{\vecphi}{\bmphi}
\safemath{\vecdelta}{\bmdelta}

\safemath{\matA}{\bA}
\safemath{\matB}{\bB}
\safemath{\matC}{\bC}
\safemath{\matD}{\bD}
\safemath{\matE}{\bE}
\safemath{\matF}{\bF}
\safemath{\matG}{\bG}
\safemath{\matH}{\bH}
\safemath{\matI}{\bI}
\safemath{\matJ}{\bJ}
\safemath{\matK}{\bK}
\safemath{\matL}{\bL}
\safemath{\matM}{\bM}
\safemath{\matN}{\bN}
\safemath{\matO}{\bO}
\safemath{\matP}{\bP}
\safemath{\matQ}{\bQ}
\safemath{\matR}{\bR}
\safemath{\matS}{\bS}
\safemath{\matT}{\bT}
\safemath{\matU}{\bU}
\safemath{\matV}{\bV}
\safemath{\matW}{\bW}
\safemath{\matX}{\bX}
\safemath{\matY}{\bY}
\safemath{\matZ}{\bZ}
\safemath{\matzero}{\bmzero}

\safemath{\matDelta}{\bDelta}
\safemath{\matLambda}{\bLambda}
\safemath{\matPhi}{\bPhi}
\safemath{\matSigma}{\bSigma}
\safemath{\matOmega}{\bOmega}
\safemath{\matTheta}{\bTheta}

\safemath{\matidentity}{\matI}
\safemath{\matone}{\matO}

\safemath{\rnda}{A}
\safemath{\rndb}{B}
\safemath{\rndc}{C}
\safemath{\rndd}{D}
\safemath{\rnde}{E}
\safemath{\rndf}{F}
\safemath{\rndg}{G}
\safemath{\rndh}{H}
\safemath{\rndi}{I}
\safemath{\rndj}{J}
\safemath{\rndk}{K}
\safemath{\rndl}{L}
\safemath{\rndm}{M}
\safemath{\rndn}{N}
\safemath{\rndo}{O}
\safemath{\rndp}{P}
\safemath{\rndq}{Q}
\safemath{\rndr}{R}
\safemath{\rnds}{S}
\safemath{\rndt}{T}
\safemath{\rndu}{U}
\safemath{\rndv}{V}
\safemath{\rndw}{W}
\safemath{\rndx}{X}
\safemath{\rndy}{Y}
\safemath{\rndz}{Z}

\safemath{\rveca}{\bimA}
\safemath{\rvecb}{\bimB}
\safemath{\rvecc}{\bimC}
\safemath{\rvecd}{\bimD}
\safemath{\rvece}{\bimE}
\safemath{\rvecf}{\bimF}
\safemath{\rvecg}{\bimG}
\safemath{\rvech}{\bimH}
\safemath{\rveci}{\bimI}
\safemath{\rvecj}{\bimJ}
\safemath{\rveck}{\bimK}
\safemath{\rvecl}{\bimL}
\safemath{\rvecm}{\bimM}
\safemath{\rvecn}{\bimN}
\safemath{\rveco}{\bomO}
\safemath{\rvecp}{\bimP}
\safemath{\rvecq}{\bimQ}
\safemath{\rvecr}{\bimR}
\safemath{\rvecs}{\bimS}
\safemath{\rvect}{\bimT}
\safemath{\rvecu}{\bimU}
\safemath{\rvecv}{\bimV}
\safemath{\rvecw}{\bimW}
\safemath{\rvecx}{\bimX}
\safemath{\rvecy}{\bimY}
\safemath{\rvecz}{\bimZ}

\safemath{\rvecxi}{\bmxi}
\safemath{\rveclambda}{\bmlambda}
\safemath{\rvecmu}{\bmmu}
\safemath{\rvectheta}{\bmtheta}
\safemath{\rvecphi}{\bmphi}

\safemath{\rmatA}{\bimA}
\safemath{\rmatB}{\bimB}
\safemath{\rmatC}{\bimC}
\safemath{\rmatD}{\bimD}
\safemath{\rmatE}{\bimE}
\safemath{\rmatF}{\bimF}
\safemath{\rmatG}{\bimG}
\safemath{\rmatH}{\bimH}
\safemath{\rmatI}{\bimI}
\safemath{\rmatJ}{\bimJ}
\safemath{\rmatK}{\bimK}
\safemath{\rmatL}{\bimL}
\safemath{\rmatM}{\bimM}
\safemath{\rmatN}{\bimN}
\safemath{\rmatO}{\bimO}
\safemath{\rmatP}{\bimP}
\safemath{\rmatQ}{\bimQ}
\safemath{\rmatR}{\bimR}
\safemath{\rmatS}{\bimS}
\safemath{\rmatT}{\bimT}
\safemath{\rmatU}{\bimU}
\safemath{\rmatV}{\bimV}
\safemath{\rmatW}{\bimW}
\safemath{\rmatX}{\bimX}
\safemath{\rmatY}{\bimY}
\safemath{\rmatZ}{\bimZ}

\safemath{\rmatDelta}{\bimDelta}
\safemath{\rmatLambda}{\bimLambda}
\safemath{\rmatPhi}{\bimPhi}
\safemath{\rmatSigma}{\bimSigma}
\safemath{\rmatOmega}{\bimOmega}
\safemath{\rmatTheta}{\bimTheta}

%% file: standard-macros.tex
\usepackage{amssymb}
\usepackage{amsfonts}
\usepackage{mathrsfs}
\usepackage{xspace}
\usepackage{bm}
\usepackage{fancyref}
\usepackage{textcomp}

\usepackage{multirow}
\usepackage{stmaryrd}

\newenvironment{textbmatrix}{	\setlength{\arraycolsep}{2.5pt}%
								\big[\begin{matrix}}{\end{matrix}\big]%
								\raisebox{0.08ex}{\vphantom{M}}}

\def\be{\begin{equation}}
\def\ee{\end{equation}}
\def\een{\nonumber \end{equation}}
\def\mat{\begin{bmatrix}}
\def\emat{\end{bmatrix}}
\def\btm{\begin{textbmatrix}}
\def\etm{\end{textbmatrix}}

\def\ba#1\ea{\begin{align}#1\end{align}}
\def\bas#1\eas{\begin{align*}#1\end{align*}}
\def\bs#1\es{\begin{split}#1\end{split}}
\def\bg#1\eg{\begin{gather}#1\end{gather}}
\def\bml#1\eml{\begin{multline}#1\end{multline}}
\def\bi#1\ei{\begin{itemize}#1\end{itemize}}

\newcommand{\lefto}{\mathopen{}\left}

\DeclareMathOperator*{\argmin}{arg\;min}		% arg min
\newcommand{\vecnorm}[1]{\lefto\lVert#1\right\rVert}		% vector norm
\safemath{\dirac}{\delta}					% Dirac delta
\safemath{\krond}{\dirac}					% Kronecker delta
\safemath{\upto}{\uparrow}
\safemath{\downto}{\downarrow}
\safemath{\iu}{j}							% imaginary unit
\safemath{\ev}{\lambda}						% eigenvalue
\safemath{\hilseqspace}{l^{2}}				% Hilbert sequence space
\newcommand{\banachfunspace}[1]{\setL^{#1}}	% Banach function space
\safemath{\hilfunspace}{\banachfunspace{2}}	% Hilbert function space
\safemath{\SNR}{\textit{SNR}} 				% signal to noise ratio
\safemath{\PAR}{\textit{PAR}} 				% signal to noise ratio
\safemath{\No}{N_0}							% noise spectral density
\safemath{\Es}{E_s}							% energy per symbol
\safemath{\Eb}{E_b}							% energy per bit
\safemath{\EbNo}{\frac{\Eb}{\No}}
\safemath{\EsNo}{\frac{\Es}{\No}}

\DeclareMathOperator{\CHop}{\ensuremath{\opH}} % channel operator
\safemath{\tvir}{\rndh_{\CHop}}				% time-varying impulse response
\safemath{\tvtf}{\rndl_{\CHop}}				% 	-''- transfer function
\safemath{\spf}{\rnds_{\CHop}}				% spreading function
\safemath{\bff}{H_{\CHop}}					% bi-freuqency function

\safemath{\ircf}{r_{h}}						% impulse response correlation fn.
\safemath{\tftvcf}{r_{s}}					% scattering function
\safemath{\tfcf}{r_{l}}						% time-frequency correlation fn.
\safemath{\bfcf}{r_{H}}						% bi-frequency correlation fn.

\safemath{\tcorr}{c_h}						% time-correlation function
\safemath{\scf}{c_{s}}						% spreading function
\safemath{\tfcorr}{c_{l}}					% transfer-function correlation
\safemath{\fcorr}{c_{H}}						% frequency-correlation function

\safemath{\mi}{I}							% mutual information
\safemath{\capacity}{C}						% capacity

\safemath{\normal}{\mathcal{N}}			% normal distribution
\safemath{\jpg}{\mathcal{CN}}			% jointly proper Gaussian
\safemath{\mchain}{\leftrightarrow}		% Markov chain
\safemath{\dB}{\,\mathrm{dB}}
\safemath{\dBm}{\,\mathrm{dBm}}
\safemath{\Hz}{\,\mathrm{Hz}}
\safemath{\kHz}{\,\mathrm{kHz}}
\safemath{\MHz}{\,\mathrm{MHz}}
\safemath{\GHz}{\,\mathrm{GHz}}
\safemath{\s}{\,\mathrm{s}}
\safemath{\ms}{\,\mathrm{ms}}
\safemath{\mus}{\,\mathrm{\text{\textmu}s}}
\safemath{\ns}{\,\mathrm{ns}}
\safemath{\ps}{\,\mathrm{ps}}
\safemath{\meter}{\,\mathrm{m}}
\safemath{\mm}{\,\mathrm{mm}}
\safemath{\cm}{\,\mathrm{cm}}
\safemath{\m}{\,\mathrm{m}}
\safemath{\W}{\,\mathrm{W}}
\safemath{\mW}{\, \mathrm{mW}}
\safemath{\J}{\,\mathrm{J}}
\safemath{\K}{\,\mathrm{K}}
\safemath{\bit}{\,\mathrm{bit}}
\safemath{\nat}{\,\mathrm{nat}}

\safemath{\define}{\triangleq}			% definition

\safemath{\equivalent}{\sim}
\safemath{\distas}{\sim}					% distributed according to
\safemath{\sdiff}{\Delta}				% symmetric set difference

\safemath{\reals}{\mathbb{R}}
\safemath{\positivereals}{\reals_{+}}
\safemath{\integers}{\mathbb{Z}}
\safemath{\posint}{\integers_{+}}
\safemath{\naturals}{\mathbb{N}}
\safemath{\posnaturals}{\naturals_{+}}
\safemath{\complexset}{\mathbb{C}}
\safemath{\rationals}{\mathbb{Q}}

\newcommand*{\fancyrefapplabelprefix}{app}		% Appendix
\newcommand*{\fancyrefthmlabelprefix}{thm}		% Theorem
\newcommand*{\fancyreflemlabelprefix}{lem}		% Lemma
\newcommand*{\fancyrefcorlabelprefix}{cor}		% Corollary
\newcommand*{\fancyrefdeflabelprefix}{def}		% Definition
\newcommand*{\fancyrefproplabelprefix}{prop}		% Proposition
\newcommand*{\fancyrefexmpllabelprefix}{exmpl}
\newcommand*{\fancyrefalglabelprefix}{alg}		% Algorithm
\newcommand*{\fancyreftbllabelprefix}{tbl}		% Algorithm

\frefformat{vario}{\fancyrefseclabelprefix}{Section~#1}
\frefformat{vario}{\fancyrefthmlabelprefix}{Theorem~#1}
\frefformat{vario}{\fancyreftbllabelprefix}{Table~#1}
\frefformat{vario}{\fancyreflemlabelprefix}{Lemma~#1}
\frefformat{vario}{\fancyrefcorlabelprefix}{Corollary~#1}
\frefformat{vario}{\fancyrefdeflabelprefix}{Definition~#1}
\frefformat{vario}{\fancyreffiglabelprefix}{Figure~#1}
\frefformat{vario}{\fancyrefapplabelprefix}{Appendix~#1}
\frefformat{vario}{\fancyrefeqlabelprefix}{(#1)}
\frefformat{vario}{\fancyrefproplabelprefix}{Proposition~#1}
\frefformat{vario}{\fancyrefexmpllabelprefix}{Example~#1}
\frefformat{vario}{\fancyrefalglabelprefix}{Algorithm~#1}

%% file: defs.tex
 \newtheorem{thm}{Theorem}
 \newtheorem{cor}[thm]{Corollary}   % Turned off theorem numbering
 
 \newtheorem{defi}{Definition}
 \newtheorem{lem}[thm]{Lemma}
 
\safemath{\dictab}{[\,\dicta\,\,\dictb\,]}

\safemath{\ysig}{\bmy}
\safemath{\ysighat}{\hat{\ysig}}
\safemath{\ysigdim}{M}
\safemath{\xsig}{\bmx}
\safemath{\xsigdim}{N}
\safemath{\nx}{n_x}
\safemath{\zsig}{\bmz}
\safemath{\zsigdim}{\ysigdim}
\safemath{\rsig}{\bmr}
\safemath{\Adict}{\bA}
\safemath{\Adicttilde}{\widetilde{\Adict}}
\safemath{\Adictdim}{\outputdim\times\xsigdim}
\safemath{\avec}{\bma}
\safemath{\avectilde}{\tilde{\avec}}
\safemath{\Bdict}{\bB}
\safemath{\Bdicttilde}{\widetilde{\Bdict}}
\safemath{\Cdict}{\bC}
\safemath{\cvec}{\bmc}
\safemath{\Ddict}{\bD}
\safemath{\Ddictdim}{\ysigdim\times\xsigdim}
\safemath{\dvec}{\bmd}
\safemath{\Ddicttilde}{\widetilde{\bD}}
\safemath{\Bonb}{\bB}
\safemath{\bvec}{\bmb}
\safemath{\Bonbdim}{\ysigdim\times\ysigdim}
\safemath{\noise}{\bmn}
\safemath{\noisedim}{\ysigim}
\safemath{\err}{\bme}
\safemath{\errdim}{\ysigdim}
\safemath{\errset}{\setE}
\safemath{\nerr}{n_e}
\safemath{\delop}{\bP_\errset}
\safemath{\delopc}{\bP_{{\errset}^c}}

\safemath{\cplxi}{\imath}
\safemath{\cplxj}{\jmath}
\safemath{\dict}{\matD}
\safemath{\inputdim}{N}		% number of columns of dictionary D
\safemath{\outputdim}{M}		%number of rows of dictionary D
\safemath{\sparsity}{S}	%sparsity
\safemath{\inputdimA}{{N_a}}	%total number of elements in dictionary A
\safemath{\inputdimB}{{N_b}}	%total number of elements in dictionary B
\safemath{\elemA}{{n_a}}	%number of elements chosen from dictionary A
\safemath{\elemB}{{n_b}}	%number of elements chosen from dictionary B
\safemath{\resA}{\matR_a}	%restriction map to elements of dictionary A
\safemath{\resB}{\matR_b}	%restriction map to elements of dictionary B
\safemath{\subD}{\matS} %subdictionary
\safemath{\subA}{\matS_a} %subdictionary part of A
\safemath{\subB}{\matS_b} %subdictionary part of B
\safemath{\dicta}{\matA} 	% first subdictionary
\safemath{\dictb}{\matB} 	% second subdictionary
\safemath{\hollowS}{H}
\safemath{\hollowA}{H_a}
\safemath{\hollowB}{H_b}
\safemath{\cross}{Z}
\safemath{\coh}{\mu_d}			% coherence dictionary
\safemath{\coha}{\mu_a}			% coherence first subdictionary
\safemath{\cohb}{\mu_b}			% coherence second subdictionary
\safemath{\mubs}{\nu}	%block sub-coherence
\safemath{\cohm}{\mu_m} %mutual coherence
\safemath{\dictset}{\setD}	% set of dictionaries
\safemath{\dictsetp}{\dictset(\coh,\coha,\cohb)}	% set of dictionaries parametrized
\safemath{\dictsetgen}{\dictset_\text{gen}}
\safemath{\dictsetgenp}{\dictsetgen(\coh)}
\safemath{\dictsetonb}{\dictset_\text{onb}}
\safemath{\dictsetonbp}{\dictsetonb(\coh)}

\safemath{\leftside}{U}
\safemath{\rightsideA}{R_a}
\safemath{\rightsideB}{R_b}

\safemath{\indexS}{\setI_S} %set of indices participating in sub-dictionary S

\safemath{\na}{n_a}			% cardinality of set of linearly independent columns of first dictionary
\safemath{\nb}{n_b}			% cardinality of set of linearly independent columns of second dictionary
\safemath{\coeffa}{p_i}	%coefficients for columns of A
\safemath{\coeffb}{q_j}	%coefficients for columns of B
\safemath{\seta}{\setP}		% set of linearly independent columns of A
\safemath{\setb}{\setQ}     % set of linearly independent columns of B
\safemath{\setw}{\setW}	%set of n largest elements of w
\safemath{\setz}{\setZ}	%set of L-n largest elements of z
\safemath{\cola}{\veca}		% generic element of the dictionary A
\safemath{\colb}{\vecb}		% generic element of the dictionary B
\safemath{\cold}{\vecd}		% generic element of the dictionary D
\safemath{\inputvec}{\vecx} 	%coefficient vector (input)
\safemath{\error}{\vece}	%error vector
\safemath{\noiseout}{\vecz} 	%noisy output vector
\safemath{\inputvecel}{x}
\safemath{\inputveca}{\vecx_a}
\safemath{\inputvecb}{\vecx_b}
\safemath{\outputvec}{\vecy}	%output of Dictionary
\safemath{\lambdamin}{\lambda_{\mathrm{min}}}
\safemath{\elltwo}{\ell_2}
\safemath{\ellone}{\ell_1}
\safemath{\ellzero}{\ell_0}
\safemath{\ellinf}{\ell_\infty}
\safemath{\ellinftilde}{\ell_{\widetilde\infty}}
\safemath{\licard}{Z(\coh,\coha,\cohb)}
\safemath{\xsol}{\hat{x}}
\safemath{\xbord}{x_b}		%Solution at the border
\safemath{\xstat}{x_s}		%Solution stationary in l0 prob
\safemath{\xstatLone}{\tilde{x}_s}
\safemath{\order}{\mathcal{O}} %order notation (big O)
\safemath{\scales}{\Theta} %scales as
\safemath{\ones}{\mathbf{1}} %all ones matrix
\safemath{\zeroes}{\mathbf{0}} %all zeroes matrix
\safemath{\thlone}{\kappa(\coh,\cohb)} %treshold l1 problem
\safemath{\constoneA}{\delta} %constant in l1 theorem to save space
\safemath{\constoneB}{\epsilon} %constant in l1 theorem to save space
\safemath{\nlarge}{L}				   %num large elements
\safemath{\sumlarge}{S_\nlarge}
\safemath{\maxlarger}{P_\nlarge}	   % maximum in Gribonval and Nielsen
\safemath{\Pzero}{\textrm{P0}}	
\safemath{\Pone}{\textrm{P1}}
\safemath{\vecfir}{\vecw}			 % \vecv element of the kernel of the dictionary, \vecv=[\vecfir \vecsec]
\safemath{\vecsec}{\vecz}
\safemath{\elvecfir}{w}              % element of vecfir
\safemath{\elvecsec}{z}				 % element of vecsec
\safemath{\nlargefir}{n}
\safemath{\normout}{\gamma}
\safemath{\auxfun}{h}
\safemath{\supp}{\textrm{supp}}%support

\safemath{\indexa}{\ell}
\safemath{\indexb}{r}
\safemath{\indexc}{i}
\safemath{\indexd}{j}

\safemath{\project}{P}%projector

%% file: 0-abstract.tex
% !TEX root = 21ASIL.tex
% DO NOT REMOVE THE ABOVE COMMENT!

\begin{abstract}
Wireless communication systems that rely on orthogonal frequency-division multiplexing (OFDM) suffer from a high peak-to-average (power) ratio (PAR), which necessitates power-inefficient radio-frequency (RF) chains to avoid an increase in error-vector magnitude (EVM) and out-of-band (OOB) emissions. The situation is further aggravated in massive multiuser (MU) multiple-input multiple-output (MIMO) systems that would require hundreds of linear RF chains.
In this paper, we present a novel approach to joint precoding and PAR reduction that builds upon a novel $\ell^p\!-\!\ell^q$-norm formulation, which is able to find minimum PAR solutions while suppressing MU interference. 
We provide a theoretical underpinning of our approach and provide simulation results for a massive MU-MIMO-OFDM system that demonstrate significant reductions in PAR at low complexity, without causing an increase in EVM or OOB emissions.
\end{abstract}

% Wireless communication systems that rely on orthogonal frequency-division multiplexing (OFDM) suffer from a high peak-to-average (power) ratio (PAR), which necessitates power-inefficient radio-frequency (RF) chains to avoid an increase in error-vector magnitude (EVM) and out-of-band (OOB) emissions. The situation is further aggravated in massive multiuser (MU) multiple-input multiple-output (MIMO) systems that would require hundreds of linear RF chains. In this paper, we present a novel approach to joint precoding and PAR reduction that builds upon a novel $\ell^p\!-\!\ell^q$-norm formulation, which is able to find minimum PAR solutions while suppressing MU interference. We provide a theoretical underpinning of our approach and provide simulation results for a massive MU-MIMO-OFDM system that demonstrate significant reductions in PAR at low complexity, without causing an increase in EVM or OOB emissions.

%% file: 1-introduction.tex
% !TEX root = 21ASIL.tex
% DO NOT REMOVE THE ABOVE COMMENT!

\section{Introduction} 
\label{sec:intro}

Massive multi-user (MU) multiple-input multiple-output (MIMO) promises improved spectral efficiency compared to that of conventional, small-scale MIMO \cite{larsson14a}. 
Communication in channels with frequency-selective fading necessitates suitable baseband processing methods that remove inter-symbol-interference~(ISI). While orthogonal frequency-division multiplexing (OFDM) \cite{NP00} is highly effective and efficient in removing ISI, the transmitted time-domain signals exhibit, in general, a large dynamic range \cite{HL05}. In order to avoid out-of-band (OOB) emissions or an increase in error-vector magnitude (EVM) caused by signal saturation and clipping in the radio-frequency (RF) chains, OFDM necessitates the use of linear (and, hence, power-inefficient) RF circuitry.   
To this end, a plethora of dynamic range-reduction methods, known as peak-to-average (power) ratio (PAR) reduction schemes, have been proposed in the literature that combat this issue \cite{rahmatallah_survey}. 

As demonstrated in~\cite{studer13a}, the excess degrees-of-freedom offered by massive MU-MIMO systems provides the unique opportunity to  \emph{jointly} perform MU interference (MUI) removal, OFDM modulation, and PAR reduction. 
Unfortunately, the joint precoding and PAR reduction (JPP) algorithm proposed in~\cite{studer13a} exhibits prohibitively high computational complexity, which prevents its deployment in practice. 
In recent years, a host of alternative JPP methods have been proposed in \cite{cha14,wang14,studer15b,guo16,zayani19,bao16,yao19,liu19,bao18}, which aim at reducing the computational complexity. However, these methods either increase EVM and OOB emissions, or aim at minimizing the transmit signal's peaks (i.e., their $\ell^\infty$-norm), but not on minimizing the actual PAR.

\subsection{Contributions}
We propose a novel JPP method for massive MU-MIMO-OFDM systems based on a novel $\ell^p\!-\!\ell^q$-norm formulation.
We provide a theoretical underpinning of the proposed approach, which shows that it is capable of finding solutions with lower PAR than the widely-used $\ell^\infty$-norm minimization approach\mbox{\cite{cha14,wang14,studer15b,guo16,zayani19,bao16,yao19,liu19,bao18}}, while also being computationally more efficient.  
We study the fundamental trade-off between the PAR and the power increase (PINC) compared to the least-squares (LS) solutions. 
Finally, we prove the efficacy of $\ell^p\!-\!\ell^q$-norm minimization for JPP in a massive MU-MIMO-OFDM system, which demonstrates that  our new formulation achieves low PAR at moderate PINC, while perfectly removing MUI without causing an increase in EVM or OOB emissions.

\subsection{Notation}

Bold lowercase and uppercase letters represent column vectors and matrices, respectively. %
We use $a_k$ for the $k$th entry of $\veca$, 
and $[\bA]_k=\veca_k$ for the $k$th column of $\bA$. 
The superscripts $(\cdot)^*$, 
$(\cdot)^T$, and $(\cdot)^H$ stand for the matrix conjugate, transpose, and Hermitian, respectively. 
The $N\times M$ all-zeros matrix is $\mathbf{0}_{N\times M}$, the $N\times N$ identity matrix is $\bI_N$, and the $N\times N$ unitary DFT matrix is $\bF_N$.
We denote the element-wise multiplication, absolute value, and $r$th power by $\circ$, $|\cdot|$, and $(\cdot)^{\circ r}$, respectively. 
The $\ell^p$-norm is given by $\vecnorm{\bma}_p=(\sum_{k} |a_k|^p )^{1/p}$, and the Frobenius norm by $\|\bA\|_F = (\sum_{i,k} |A_{i,k}|^2)^{1/2}$.
All complex-valued gradients follow the definitions of~\cite{kreutzdelgado2009complex}. 

%% file: 2-prereqs.tex
% !TEX root = 21ASIL.tex
% DO NOT REMOVE THE ABOVE COMMENT!

\section{Prerequisites}
\label{sec:prereq}
 
We now introduce the PAR minimization problem for the simple case of $\bmy=\bA\bmx$ and study the limits of existing algorithms that find solutions $\bmx$ with low (or minimal) PAR. 
 
\subsection{Minimum PAR Solutions}
We are interested in solving an underdetermined system of linear equations $\bmy=\bA\bmx$, where $\bmy\in\complexset^M$ and $\bA\in\complexset^{M\times N}$ with $M< N$. 
While the least-squares (LS) solution vector
\leqnomode
\begin{align} \label{eq:LS}
\qquad\qquad\hat\bmx^\text{LS} = \argmin_{\tilde\bmx\in\complexset^N}\, \|\tilde\bmx\|_2 \quad \text{subject to } \bmy=\bA\tilde\bmx \tag{P-LS}
\end{align}
\reqnomode
minimizes the power in terms of the $\ell^2$-norm, we are interested in solution vectors with low dynamic range. In communication scenarios, one is typically interested in solutions with low peak-to-average (power) ratio (PAR), which is defined as follows: 
\begin{defi}
The peak-to-average-power ratio (PAR) of a non-zero vector $\bmx\in\complexset^N$ is defined as 
\begin{align} \label{eq:defPAR}
\textit{PAR}(\bmx) \define \frac{N\|\bmx\|^2_\infty}{\|\bmx\|^2_2}.
\end{align}
\end{defi}

The PAR satisfies 
$1\leq \textit{PAR}(\bmx) \leq N$,
where the upper bound  is achieved for sparse vectors with only one nonzero entry (i.e., one-sparse vectors), and the lower bound is achieved for minimum-PAR  vectors that satisfy the following definition:
\begin{defi}
A minimum PAR (min-PAR) vector $\bmx\in\complexset^N$ satisfies $|x_i|=|x_j|$ for all $i,j\in\{1,\ldots,N\}$.
\end{defi}

It would be natural to directly minimize the PAR in~\fref{eq:defPAR} of vectors $\bmx\in\complexset^N$, subject to the consistency constraint $\bmy=\bA\bmx$. Unfortunately, solving the minimum-PAR optimization problem
\leqnomode
\begin{align} \label{eq:PARminimization}
\qquad \hat\bmx^\text{MP} = \argmin_{\tilde\bmx\in\complexset^N}\, \textit{PAR}(\tilde\bmx) \quad \text{subject to } \bmy=\bA\tilde\bmx   \!\!\!\!\!\!\!\!\!\!\!\!\!\!\!\!\!\!\! 
\tag{P-PAR}
\end{align}
\reqnomode
is challenging  because the PAR in \fref{eq:defPAR} is  nonconvex and not differentiable. 
In order to efficiently solve such optimization problems, it would be beneficial to have (i) a convex problem and (ii) an objective function that is differentiable. 
While the former would enable one to find optimal solutions, the latter would allow for the design of computationally efficient algorithms, such as projected gradient descent \cite{parikh14a,BT09,GSB14}.

%%%
\subsection{Low-PAR Solutions via $\ell^\infty$-Norm Minimization}
Instead of directly minimizing the PAR, the references~\cite{studer13a,studer15b} proposed to minimize the vector's $\ell^\infty$-norm, resulting in the following  optimization problem:
\leqnomode
\begin{align} \label{eq:linftyminimization}
\qquad\quad\,\,\,\hat\bmx^\infty = \argmin_{\tilde\bmx\in\complexset^N}\, \|\tilde\bmx\|_\infty \quad \text{subject to } \bmy=\bA\tilde\bmx. \tag{P-$\infty$}
\end{align}
\reqnomode
While this problem is convex, the objective function is not differentiable, which requires Douglas-Rachford splitting~\cite{EB92} (or related splitting methods) to solve it numerically. For example, convex reduction of amplitudes for Parseval frames (CRAMP)~\cite{studer15b} is an efficient method for solving  \fref{eq:linftyminimization}. 
 
Minimizing the $\ell^\infty$-norm has shown to reduce the PAR, but the $\ell^\infty$-norm objective only focuses on reducing the signal's peaks and not on actually delivering a min-PAR solution. 
In fact, as shown in \cite[Lem.~1]{studer15b} for \fref{eq:linftyminimization} with full-spark frames\footnote{A full spark frame $\bA\in\complexset^{M\times N}$ is an $M\leq N$ matrix for which every size-$M$ subcollection of column vectors is a spanning set (i.e., has full rank).}~$\bA$, any solution to \fref{eq:linftyminimization}  is guaranteed to have $N-M+1$ entries with magnitude $\|\hat\bmx^\infty\|_\infty$, whereas no guarantee can be given for the other $M-1$ entries. This implies that the PAR of the solutions to~\fref{eq:linftyminimization} are bounded by \cite[Thm.~7]{studer15b}
\begin{align}
\textit{PAR}(\hat\bmx^\infty)  \leq \frac{N}{N-M+1}.
\end{align}
However, in order to find min-PAR solutions, other problem formulations are necessary. In \fref{sec:lplqminimization}, we will propose a novel approach that addresses this issue.

\subsection{Fundamental PAR vs.\ PINC Trade-off}
\label{sec:parpinctradeoff}
When computing solutions with low PAR, the power (or squared $\ell^2$-norm) of the solution vector typically increases compared to the LS solution (which is, per definition, power-minimal); this has the implication that, while we can shape the vector~$\bmx$ to be better suited for nonlinear RF circuitry, the resulting vector might have larger $\ell^2$-norm. Hence, for a given a power constraint, one has to back-off compared to the power of the LS solution, which will lower the SNR at the UE sides resulting a higher error-rate or lower spectral efficiency. 

We now rigorously show that  there exists a fundamental trade-off between the PAR of a solution vector $\bmx$ for the system of linear equations $\bmy=\bA\bmx$ and the power increase compared to the LS solution, which we define as follows:

\begin{defi}
Let $\bmx$ be any vector to $\bmy=\bA\bmx$ and $\hat\bmx^\text{LS}$ be the LS solution from \fref{eq:LS}. Then, the power increase (PINC) is defined as follows: 
\begin{align}
\textit{PINC}(\bmx) \define \frac{\|\bmx\|_2^2}{\|\hat\bmx^\text{LS}\|_2^2}.
\end{align}
\end{defi}

The PINC satisfies $1\leq \textit{PINC}(\bmx)$ and the lower bound is, per definition, achieved for the LS solution $\hat\bmx^\text{LS}$.
The following result reveals the fundamental trade-off between PAR and PINC; a short proof is given in \fref{app:PARPINCtradeoff}.
\begin{lem} \label{lem:PARPINCtradeoff}
Let $\bmx$ be any solution to $\bmy=\bA\bmx$ for a given $\bA$ and $\bmy$ with $\bmy\neq\bZero$. Then, there exists a constant $c\geq1$ that depends on~$\bmy$ and~$\bA$ for which the following inequality holds:
\begin{align}
\textit{PAR}(\bmx) \, \textit{PINC}(\bmx) \geq c.
\end{align}
Furthermore, the solution $\hat\bmx^\infty$ from~\fref{eq:linftyminimization} achieves the lower bound with equality, i.e., determines the constant $c$. 
\end{lem}

This result implies that, while the  $\ell^\infty$-norm solution $\hat\bmx^\infty$ is trade-off optimal, it does not necessarily minimize the PAR. It also follows that  min-PAR solutions typically increase the PINC, which is made explicit by the following corollary: 

\begin{cor}
Let $\bmy=\bA\hat\bmx$ with a min-PAR solution $\hat\bmx$. Then, its PINC is lower bounded by $c$, i.e., $\textit{PINC}(\hat\bmx)\geq c$.
\end{cor}

Our prime goal is thus  to design an optimization problem that (i)  leads to min-PAR solutions which do not result in too high PINC and (ii) enables the design of computationally efficient algorithms---this is exactly what we will do next! 

%% file: 3ab-problem.tex
% !TEX root = 21ASIL.tex
% DO NOT REMOVE THE ABOVE COMMENT!

\section{$\ell^p\!-\!\ell^q$-Norm Minimization}
\label{sec:lplqminimization}

We now introduce a novel approach to efficiently compute min-PAR solutions. We show our idea for the basic case of $\bmy=\bA\bmx$. An application to joint precoding and PAR reduction in the massive MU-MIMO case is given in~\fref{sec:ofdm}.

\subsection{$\ell^p\!-\!\ell^q$-Norm Minimization Problem}
We start by stating the following key equivalence property between $\ell^p$ and $\ell^q$ norms with $1\leq q < p$ \cite{golub}:
\begin{align} \label{eq:normequivalence}
\|\bmx\|_q \leq N^{\frac{1}{q}-\frac{1}{p}} \|\bmx\|_p.
\end{align}
Here, it is important to realize that the equality holds only for min-PAR vectors $\bmx$.
This equivalence also implies that  
\begin{align} 
%\|\bmx\|_q^2 & \leq N^{\frac{2}{q}-\frac{2}{p}} \|\bmx\|_p^2 \\
0 &  \leq N^{\frac{2}{q}-\frac{2}{p}} \|\bmx\|_p^2 - \|\bmx\|_q^2,
\end{align}
where the last inequality holds, once again, only for min-PAR vectors $\bmx$. 
Our key idea is to  minimize the right-hand-side, resulting in the following optimization problem for $1\leq q < p$:
\begin{align} \label{eq:lplqminimization}
\!\!\!\hat\bmx = \argmin_{\tilde\bmx\in\complexset^N} N^{\frac{2}{q}-\frac{2}{p}} \|\tilde\bmx\|_p^2 \!-\! \|\tilde\bmx\|_q^2 \,\  \,\text{subject to } \bmy=\bA\tilde\bmx.\!\! \tag{P-$pq$}
\end{align}
This problem is nonconvex, but the differentiability of the objective function (assuming $p<\infty$) enables computationally efficient algorithms. 
Furthermore, the objective is minimal only for min-PAR solution vectors $\hat\bmx$, which is our prime goal.

%%%

\subsection{Why the $\ell^p\!-\!\ell^q$-Norm?}
It is natural to ask, why the objective function of~\fref{eq:lplqminimization}, which we define as
\begin{align} \label{eq:objf}
f(\bmx) \define  N^{\frac{2}{q}-\frac{2}{p}} \|\bmx\|_p^2 \!-\! \|\bmx\|_q^2
\end{align}  
is a sensible choice for minimizing the PAR. Our reasons are as follows. 
First, we can show that minimizing~$f(\bmx)$ is, under certain conditions, equivalent to minimizing the following, alternative  $\ell^p\!-\!\ell^q$-norm-based PAR definition:
\begin{defi} 
For $1\leq q < p$, the $\ell^p\!-\!\ell^q$-norm-based PAR of a non-zero vector $\bmx\in\complexset^N$  is defined as 
\begin{align} \label{eq:defPARpq}
\PARpq{p}{q}(\bmx) \define \frac{N^{\frac{2}{q}-\frac{2}{p}}\|\bmx\|^2_p}{\|\bmx\|^2_q}.
\end{align}
\end{defi}

This PAR definition satisfies $1 \leq \PARpq{p}{q}(\bmx) \leq N^{\frac{2}{q}-\frac{2}{p}}$, where the upper and lower bounds are achieved by  one-sparse and min-PAR vectors, respectively, similarly to the standard PAR in~\fref{eq:defPAR}.
In addition, we have the following inequality; a short proof is given in  \fref{app:parpqinequality}.
\begin{lem} \label{lem:parpqinequality}
For any nonzero vector $\bmx\in\complexset^N$, the $\text{PAR}_q^p$ satisfies 
\begin{align}
\PARpq{p}{q}(\bmx) \leq \PAR(\bmx),
\end{align}
for (i) any $2 \leq q < p$, and for (ii) $q=1$ and $p=2$. 
This bound holds with equality for min-PAR solutions, and also trivially for $q=2$ and $p=\infty$.
\end{lem}
Now, consider the following optimization problem:
\begin{align} \label{eq:PARpqminimization}
\hat\bmx^\star = \argmin_{\tilde\bmx\in\complexset^N}\, \PARpq{p}{q}(\bmx)  \quad \text{subject to } \bmy=\bA\tilde\bmx.
\end{align}
By following the arguments in~\cite{dinkelbach1967nonlinear}, it can be shown that as long as a min-PAR solution exists, the solution to~\fref{eq:PARpqminimization} is the same as the solution to \fref{eq:lplqminimization}.
These facts imply that minimizing the alternative definition of~$\textit{PAR}_p^{q}$ is a viable substitute to minimizing $\textit{PAR}$ in~\fref{eq:defPAR}.
%

%% file: 3cd-solution.tex
% !TEX root = 21ASIL.tex
% DO NOT REMOVE THE ABOVE COMMENT!

\subsection{$\ell^p\!-\!\ell^q$-Norm Minimization with Forward Backward Splitting}
\label{sec:lplqviaFBSisfun}
The remaining ingredient is a computationally efficient method to solve \fref{eq:lplqminimization}.
Unfortunately, the nonconvex nature of~\fref{eq:lplqminimization} makes finding global minimizers difficult. Nonetheless, we next develop an algorithm that requires low complexity and is shown to find min-PAR solutions.  
Specifically, we will use forward backward splitting (FBS)~\cite{BT09}, a numerical optimization procedure that solves problems of the form
\begin{align} \label{eq:generalproblem}
\hat{\bmx} = \argmin_{\bmx\in\complexset^N} f(\bmx)+g(\bmx),
\end{align}
where the function $f$ is convex and differentiable and $g$ is convex, but not necessarily smooth or bounded.
FBS solves~\fref{eq:generalproblem} by performing the following iterative procedure for the iterates $k=1,2,\ldots$, until a convergence criterion is met:
\begin{align}\label{eq:FBS_iterations}
\bmx^{(k+1)} = \text{prox}_g( \bmx^{(k)} - \tau^{\revision{(k)}} \  \nabla f(\bmx^{(k)}), \tau^{(k)}).
\end{align}
Here, the proximal operator is defined as %\cite{BT09}
\begin{align} \label{eq:proximal}
\text{prox}_g(\vecz,\tau^{(k)}) = \argmin_{\bmx\in\complexset^N} \,\bigl\{\tau^{(k)} g(\vecx) + \textstyle \frac{1}{2}\|\vecx-\vecz\|_2^2\bigr\}
\end{align}
with the per-iteration step size $\tau^{\revision{(k)}}>0$ and $\nabla f$ is the gradient of the smooth function $f$.
FBS is guaranteed to converge to a solution of \fref{eq:generalproblem} for carefully-chosen step sizes. %~\cite{GSB14}. 
It is important to note that FBS has been used in the past to efficiently approximate solutions to problems of the form~\fref{eq:generalproblem} in which~$f$ is no longer convex~\cite{liang2016multi,boct2016inertial}, as it is the situation for~\fref{eq:lplqminimization}. 

In our case of applying FBS to solving~\fref{eq:lplqminimization}, the function~$f(x)$ is defined in~\fref{eq:objf} and its gradient is given by 
\begin{align} \label{eq:gradient}
& \!\!\nabla \!f(\bmx) = (N^{\frac{2}{q}-\frac{2}{p}} \|\bmx\|_p^{2-p} |\bmx|^{p-2}\!  - \|\bmx\|_q^{2-q}|\bmx|^{q-2})\!\circ\! \bmx.
\end{align}
The function $g(\bmx)$ is used to represent the constraint of~\fref{eq:lplqminimization}, which can be accomplished by setting $g(\bmx)$ 
to a characteristic function  that is zero if $\bmy=\bA\bmx$ and infinity otherwise.
For this case, the proximal operator is given by~\cite{GSB14}
\begin{align} \label{eq:proximal}
\text{prox}_g(\vecz) = \bmz - \bA^H(\bA\bA^H)^{-1}(\bA\vecz-\vecy),
\end{align}
which is independent of $\tau^{(k)}$ and requires $\bA\bA^H$ to be invertible.  
In what follows, we pick a fixed step size $\tau=\tau^{(k)}$ that results in a  monotonic decrease in the objective. 
We use FBS as in~\fref{eq:FBS_iterations} for a fixed maximum number of iterations $K_\text{max}$.

\begin{figure}[tp]
\centering
\vspace{-0.2cm}
\subfigure[PINC vs.\ PAR trade-off.]{\includegraphics[width=0.49\columnwidth]{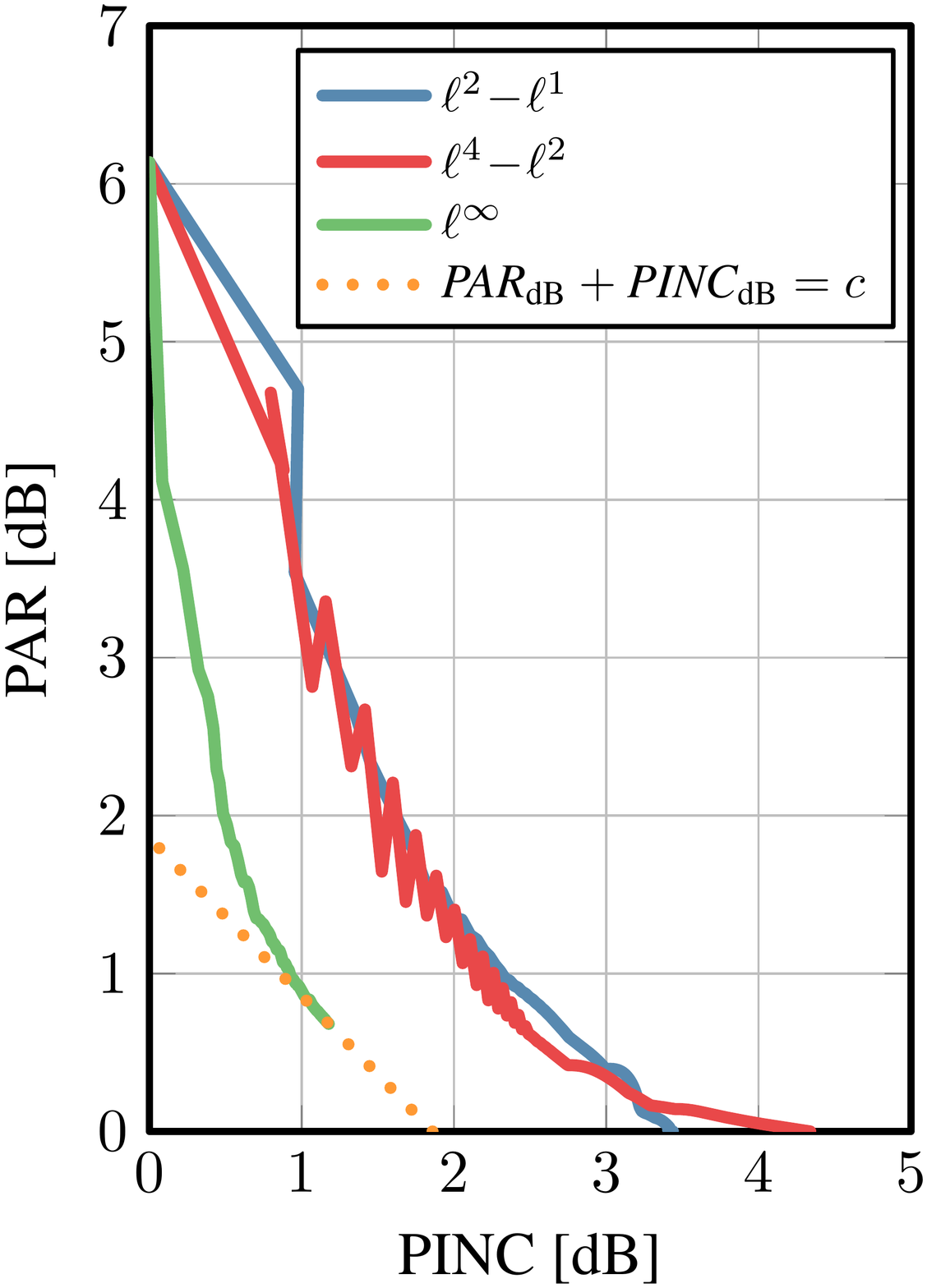}\label{fig:simpletradeoff}}
\subfigure[Convergence of PINC and PAR.]{\includegraphics[width=0.49\columnwidth]{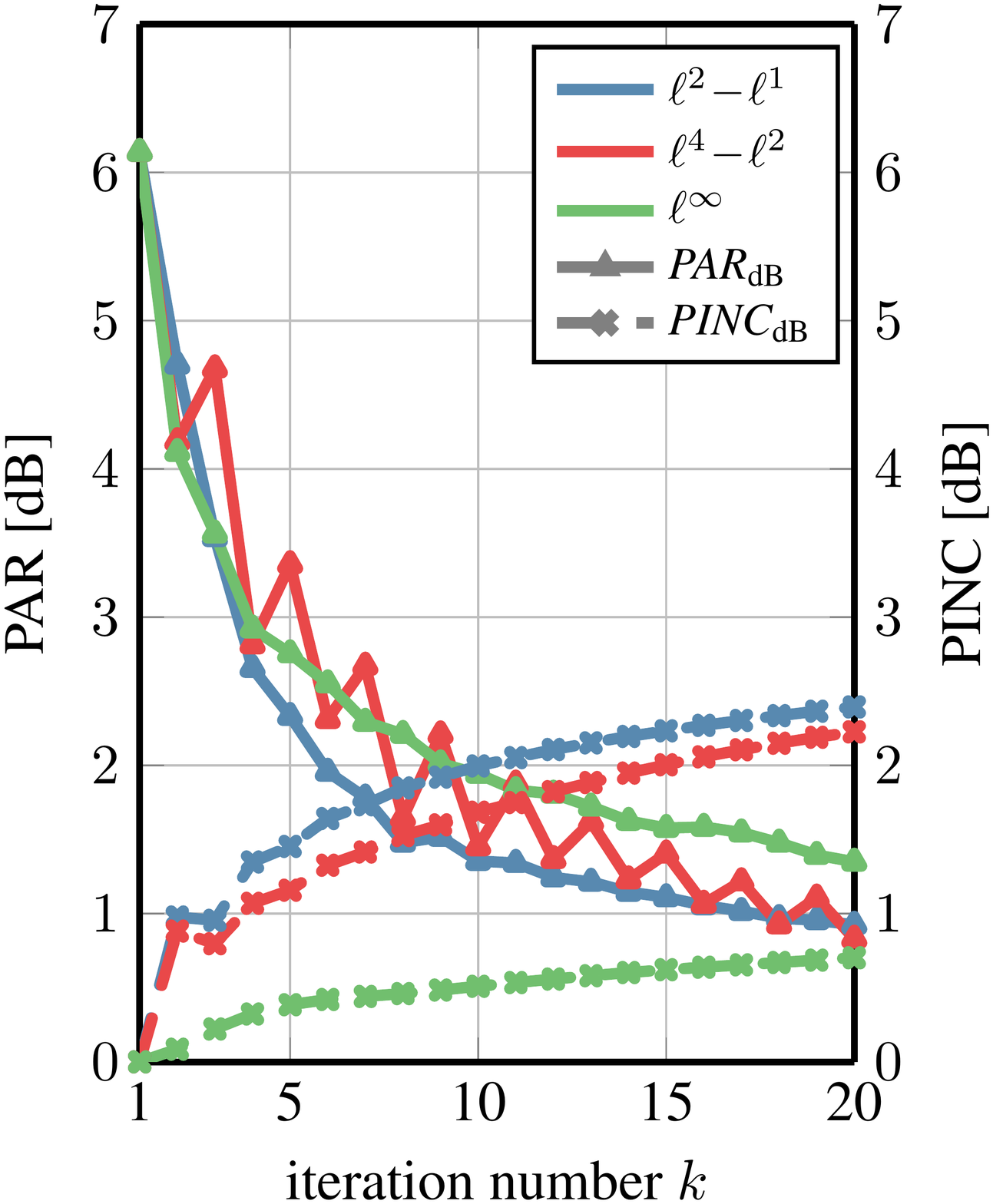}\label{fig:simpleiter}}
\caption{Comparison between $\ell^p\!-\!\ell^q$-norm and $\ell^\infty$-norm minimization for a circularly-symmetric complex standard normal matrix  $\bA\in\opC^{100\times 200}$ and vector $\vecy\in\opC^{100}$: (a) Trade-off between PAR and PINC; (b) PAR (solid lines) and PINC (dashed lines) for the first 20  iterations. The $\ell^p\!-\!\ell^q$-norm-based methods solved with FBS are able to compute min-PAR solutions. }
 \label{fig:simplecase}
\end{figure}

\subsection{Example of $\ell^p\!-\!\ell^q$-Norm Minimization}
\label{sec:simpleex}

In \fref{fig:simplecase}, we show an example of $\ell^p\!-\!\ell^q$-norm minimization, where we apply FBS to one instance of a circularly-symmetric complex standard normal $\bA\in\opC^{100\times 200}$ and vector $\vecy\in\opC^{100}$.
We consider the cases where $p=4,q=2$ and $p=2,q=1$; moreover, as a baseline, we show the behavior of $\ell^\infty$-norm minimization solved via  CRAMP~\cite{studer15b}.
For all algorithms, we start with $\bmx^{(1)}=\text{prox}_g(\mathbf{0}_{100\times 1})=\hat\bmx^\text{LS}$, i.e., all algorithms compute the LS solution in the first iteration, and run a maximum of $K_\text{max}=10^6$ iterations.

\fref{fig:simpletradeoff} shows the PAR-PINC trade-off for all iterations together with the lower-bound given by \fref{lem:PARPINCtradeoff}:
\begin{align}
\textit{PAR}_\text{dB}(\bmx) +\textit{PINC}_\text{dB}(\bmx) \geq 10\log_{10}(c).
\end{align}
While the minimum $\ell^\infty$-norm solution indeed achieves the optimal trade-off, the $\ell^p\!-\!\ell^q$-norm variants actually compute min-PAR solutions---but, as expected, at higher PINC.  
We also note the non-monotone behavior of the PAR for the $\ell^p\!-\!\ell^q$-norm-based algorithms, as they only guarantee a monotonic decrease for their own objective  in \fref{eq:objf}.
\fref{fig:simpleiter} shows the behavior of PAR and PINC for the first 20 iterations, which is more realistic in practical systems that are limited in complexity and latency. 
We observe that the per-iteration behavior is comparable. 
However, the proposed $\ell^p\!-\!\ell^q$-norm methods are more efficient since Douglas-Rachford splitting requires more intermediate variables (and, hence, more storage) and also a higher per-iteration complexity, mostly because evaluating the proximal operator for the $\ell^\infty$-norm~\cite{studer15b} is significantly more complex than evaluating the gradient in~\fref{eq:gradient}.

%% file: 4-ofdm.tex
% !TEX root = 21ASIL.tex
% DO NOT REMOVE THE ABOVE COMMENT!
\section{The Massive MU-MIMO-OFDM Case}
\label{sec:ofdm}

We now apply $\ell^p\!-\!\ell^q$-norm minimization to joint precoding and PAR reduction in a massive MU-MIMO-OFDM system.

\subsection{System Model}
\label{sec:sys_model}

\begin{figure}[tp]
\centering
\vspace{0.1cm}
\includegraphics[width=.99\columnwidth]{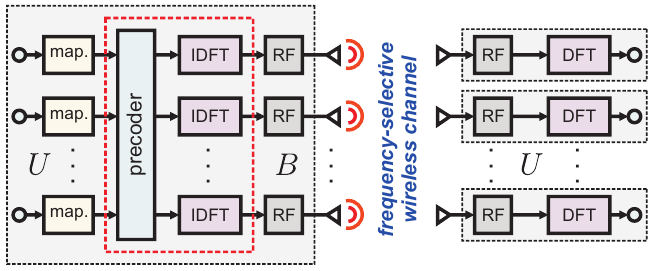}
\vspace{-0.3cm}
\caption{Massive MU-MIMO-OFDM system overview. The proposed $\ell^p\!-\!\ell^q$-norm joint precoding and PAR reduction (JPP) method is located at the BS side (left) and highlighted with a  dashed red box.} 
\label{fig:lame_figure}
\end{figure}

We consider a massive MU-MIMO-OFDM downlink system as depicted in \fref{fig:lame_figure}. A BS equipped with $B$ antennas transmits data to~$U<B$ single-antenna UEs.
We assume that the total number of OFDM tones is $W$, and we designate the sets of used and unused OFDM tones with $\Omega$ and $\Omega^c$, respectively, where $|\Omega|+|\Omega^c|=W$.
For a given OFDM tone $w\in\Omega$, the signal vector $\vecs_w\in\mathcal{S}^U$ to be transmitted contains the data symbols from the constellation $\mathcal{S}$ for each UE; we set $\vecs_w=\mathbf{0}_{U\times 1}$ for the unused tones $w\in\Omega^c$.
In order to suppress MUI, the signal vectors $\vecs_w,w\in\Omega$ are passed through a precoder that generates $W$ frequency-domain vectors $\vecx_w\in\opC^B$ according to a given precoding scheme.
We define the matrix $\bX\in\opC^{B\times W}$ so that the columns consist of the precoded vectors $\vecx_w,w=1,\dots,W$, i.e., $\bX=[\bmx_1,\ldots,\bmx_W]$, where each column corresponds to a tone and each row to a BS antenna.
Then, the transpose $\bX^T$ provides the frequency domain outputs for each BS antenna, i.e., each column corresponds to a BS antenna and each row  to a tone. 
Since precoding causes the total transmit power $P =\|\bX\|_F^2$ to depend on the transmit signals $\vecs_w$, $\forall w$, and the channel state, the precoded vectors will be  normalized prior to transmission as $\hat\vecx_w = \vecx_w /\|\bX\|_F$ to ensure unit transmit power.
This normalization is essential in practice (i.e., to meet regulatory power constraints), but we omit this normalization step in the description of the precoders to follow (but we will recall this aspect in~\fref{sec:results}).

Let us denote the time-domain output samples of the $B$ BS antennas by $\{\vect_b\}_{b=1}^B$ and define the $W\times B$ matrix $\bT=[\bmt_1,\ldots,\bmt_B]$. In an OFDM system, the matrix $\bT$ is given by the inverse DFT as $\bT = \bF^H\bX^T$.
To simplify notation, we define a linear mapping from $\{\vect_b\}_{b=1}^B$ to $\{\vecx_w\}_{w=1}^{W}$ as
\begin{align} \label{eq:psithemapper}
\psi_w(\vect_1,\dots,\vect_b) \define  [(\bF\bT)^T]_w=\vecx_w,\, w\in\{1,\dots,W\}.
\end{align}
Prior to transmission over the wireless channel, a cyclic prefix (CP) is prepended to $\vect_b,\forall b\in \{1,\dots,B\}$ to avoid ISI.

For simplicity, we specify the input-output relation of the wireless channel in the frequency domain. We model the received vector at tone index $w$ with $\vecy_w\in\opC^W$ as
\begin{align} \label{eq:y_Hs}
    \vecy_w = \bH_w \vecx_w + \vecn_w,\, w\in\{1,\dots,W\}.
\end{align}
where $\bH_w\in\opC^{U\times B}$ represents the MIMO channel matrix associated with the $w$th OFDM tone and $\vecn_w\in\opC^U$ models circularly-symmetric Gaussian noise.
Finally, each of the $U$ UEs perform OFDM demodulation to obtain $[\vecy_w]_u, w \in\Omega$, i.e., the data symbols for UE $u$ at each used tone.

\subsection{Least-Squares Precoding}
\label{sec:LSprecoding}
In order to suppress MUI, precoding must be employed at the BS. To this end, we assume the channel matrices $\bH_w, \forall w$ to be
known perfectly at the BS-side\footnote{In massive MU-MIMO systems, channel-state information can be acquired through pilot-based training in the uplink and by exploiting reciprocity.}. 
Linear precoders are among the simplest methods and compute $\vecx_w = \bG_w \vecs_w$, with the precoding matrix $\bG_w\in\opC^{B\times U}$, on the used subcarriers $w\in\Omega$ and $\bmx_w=\bZero_{B\times 1}$ on the unused subcarriers $w\in\Omega^c$.
LS precoding is a prominent method, which determines the vectors~$\bmx_w$, $\forall w$, so that  the following precoding constraints are satisfied while minimizing the transmit power: (i) $\vecs_w = \bH_w\vecx_w$, $w\in\Omega$, which ensure zero EVM, and (ii) $\bmx_w=\bZero_{B\times 1}$, $w\in\Omega^c$, which ensure zero OOB omissions. 
The precoding constraints~(i) have a known closed-form solution with the precoding matrices being $\bG_w=\bH_w^H (\bH_w\bH_w^H)^{-1}$, $w\in\Omega$.

While LS precoding perfectly eliminates MUI and results in minimal PINC, the PAR of the resulting time-domain signals is typically very high~\cite{studer13a,cha14,wang14,studer15b,guo16,zayani19,bao16,yao19,liu19,bao18} which would require highly-linear (and, hence, power inefficient and costly) RF circuitry.

\subsection{$\ell^p\!-\!\ell^q$-Norm Joint Precoding and PAR Reduction}
\label{sec:precoding}
Inspired by the work in \cite{studer13a}, massive MU-MIMO has the unique property that the downlink channel has a large nullspace, which can be exploited to simultaneously satisfy the precoding constraints (which removes MUI), while shaping the transmitted time-domain signals to reduce the PAR.
Consequently, our goal is to solve an optimization problem that simultaneously satisfies the following precoding constraints
while minimizing the $\ell^p\!-\!\ell^q$-norm of the time-domain signals: (i) $\vecs_w = \bH_w\vecx_w$, $w\in\Omega$, which ensure zero EVM, and (ii) $\bmx_w=\bZero_{1\times B}$, $w\in\Omega^c$, which ensure zero OOB emissions. 
We propose the following optimization problem that achieves all of these goals:
\begin{align*} 
(\text{JPP-}pq)\!\! \,\, \left\{\begin{array}{cl} 
\underset{\vect_1,\dots,\vect_B \in \complexset^W}{\text{minimize}} & \displaystyle  \sum_{b=1}^B\big( W^{\frac{2}{q}-\frac{2}{p}}\|\vect_b\|_p^2 \!-\! \|\vect_b\|_q^2 \big) \\[0.4cm]
\text{subject to} & \vecs_w = \bH_w\psi_w(\vect_1,\dots,\vect_b) , \forall w\in\Omega\\[0.1cm]
& \mathbf{0}_{U\times 1} = \psi_w(\vect_1,\dots,\vect_b) ,\, \forall w\in\Omega^c.
\end{array}\right.
\end{align*}
Here, we decided to separately minimize $\text{PAR}_q^p$ at each transmit antenna and consider the sum of $\ell^p\!-\!\ell^q$-norm penalties.\footnote{Another approach would be  to minimize $\text{PAR}_q^p$ over all time-domain signals, i.e., $\text{PAR}_q^p(\text{vec}(\bT))$. A detailed investigation of this alternative objective function is part of future work.}

The remaining piece of the puzzle is to show that a solution to $(\text{JPP-}pq)$ can be computed efficiently (and approximately) via FBS. In fact, the procedure is just a slightly more complicated version of the algorithm proposed in \fref{sec:lplqviaFBSisfun} and the details are as follows. Following the definition of $f$ in \fref{eq:objf},
the objective function is given by $\tilde f(\bT)  \define \sum_{b=1}^B f(\vect_b)$. 
Here, the summation allows the computation of the gradient step in $\tilde f(\bT)$ separately for each column  $\vect_b$, $b\in\{1,\dots,B\}$.
Since there exists a one-to-one mapping between time and frequency domains via~\fref{eq:psithemapper},
we can apply the linear constraints in $(\text{JPP-}pq) $ separately on the columns of the frequency domain matrix $\bX$. Here, we apply \fref{eq:proximal} on $\vecx_w$ for $w\in\Omega$, and set $\vecx_w=\mathbf{0}_{U\times 1}$ for $w\in\Omega^c$.
We then repeat the resulting FBS procedure for a fixed (and small) number of iterations $K_\text{max}$.

%% file: 5-results.tex
% !TEX root = 21ASIL.tex
% DO NOT REMOVE THE ABOVE COMMENT!

\section{Simulation Results}
\label{sec:results}

We now demonstrate the efficacy of $\ell^p\!-\!\ell^q$-norm minimization for JPP in a massive MU-MIMO-OFDM system. 

\subsection{Simulation Setup}
As in \cite{jacobsson17f}, we consider a MU-MIMO-OFDM system with $B=128$ BS antennas and $U=16$ UEs.
The OFDM numerology is based on 20\,MHz bandwidth with $W=2048$ subcarriers; the used and unused tones are as defined in \cite{3gpp19a}.
We assume a Rayleigh fading channel model with $L=4$ taps, where the entries of the non-zero time-domain matrices $\bH_t , t = 1,\dots, L$ are assumed i.i.d.\ circularly complex Gaussian with unit variance. The frequency domain channel matrices are obtained using the Fourier transform  as detailed in~\cite{studer16a}.
The transmit data symbols are taken from a 16-QAM constellation. 
As mentioned in \fref{sec:sys_model}, the precoded vectors have to be normalized to unit power before transmission.
This back-off in the transmit power is equivalent to an SNR decrease at the UE side; in other words, the PINC directly translates to an SNR performance loss of exactly $\textit{PINC}_\text{dB}$. 

We solve  $(\text{JPP-}pq)$ as explained in \fref{sec:precoding}.
As in \fref{sec:simpleex}, we consider the cases $p=4,q=2$ and $p=2,q=1$. As a baseline, we also compare with $\ell^\infty$-norm minimization solved using CRAMP~\cite{studer15b}.
All algorithms produce the LS solution  in the first iteration, i.e., $\bmx_w^{(1)}=\hat\bmx_w^\text{LS}, w\in\Omega$, $\bmx_w^{(1)}=\bZero_{B\times 1}, w\in\Omega^c$, and run for  $K_\text{max}=20$ iterations.
We reiterate that all of these algorithms do not increase the EVM or cause any OOB emissions; this means that the resulting error-rate performance is equivalent to that of the LS precoder up to an SNR gap that is determined solely by the PINC.

\begin{figure}[tp]
\centering
\includegraphics[width=.98\columnwidth]{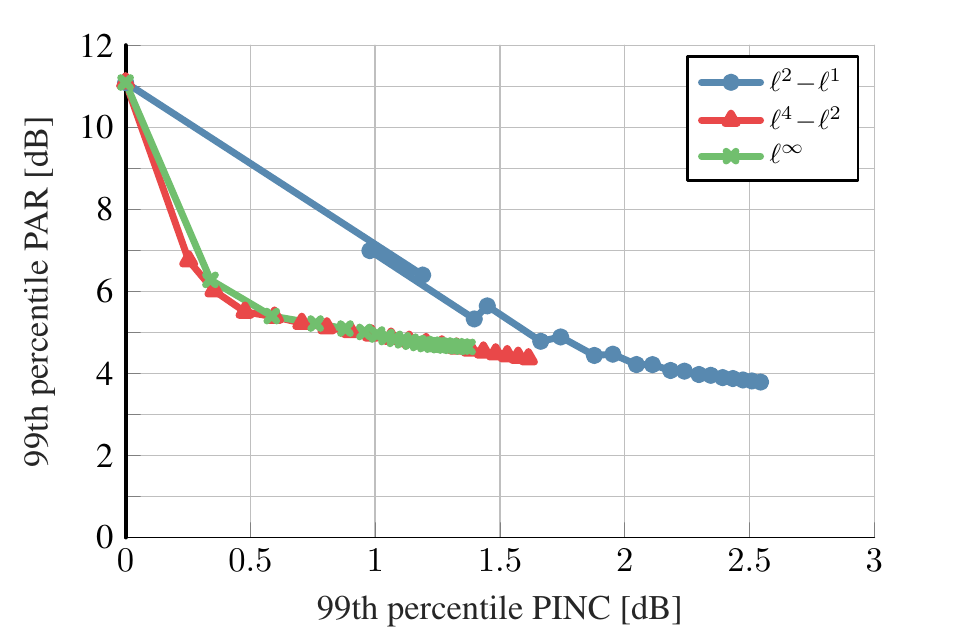}
\vspace{-0.3cm}
\caption{PAR vs.\ PINC trade-off for $\ell^p\!-\!\ell^q$-norm-based JPP methods in a massive MU-MIMO-OFDM system. The markers correspond to iterations and all algorithms start with the LS solution at $11.1$\,dB PAR and $0$\,dB PINC.}
\label{fig:pinc_par_tradeoff}
%\vspace{-0.3cm}
\end{figure}

\subsection{Results and Discussion}
\fref{fig:pinc_par_tradeoff} shows the trade-off between PINC and PAR for JPP in this massive MU-MIMO-OFDM scenario.
 We use Monte-Carlo sampling to compute the complementary cumulative distribution function (CCDF) for the PAR and PINC at each iteration for the different JPP methods.
The CCDF is defined as
$\text{CCDF}_Z(z) = \mathbb{P}(Z > z)$ for a random variable $Z$. Here, the value $z$ for which $\text{CCDF}_Z(z) = 1$\% is the 99th percentile of~$Z$. 
Following this definition, we pick the 99th percentile as the operating point for both $\textit{PAR}_{\text{dB}}$ and $\textit{PINC}_{\text{dB}}$, and we show the PAR-PINC trade-off for each algorithm iteration in \fref{fig:pinc_par_tradeoff}, where the iterations are designated by the markers.
Similar to \fref{sec:simpleex}, we observe that the $\ell^p\!-\!\ell^q$-norm methods are able to compute lower-PAR solutions than $\ell^\infty$-norm minimization, but at higher PINC. 
We also note that all three algorithms decrease the PAR by approximately 5\,dB in only one iteration. 
Since the $\ell^p\!-\!\ell^q$-norm methods are more efficient than $\ell^\infty$-norm minimization for the reasons discussed in \fref{sec:simpleex}, our proposed approach turns into a significant complexity advantage for a $B\times U\times W$-sized massive MU-MIMO-OFDM system.

%% file: 6-conclusion.tex
% !TEX root = 21ASIL.tex
% DO NOT REMOVE THE ABOVE COMMENT!

\section{Conclusions}

We have proposed a novel formulation for finding minimal-PAR solutions to underdetermined systems of linear equations using $\ell^p\!-\!\ell^q$-norm minimization.
We have identified a fundamental trade-off between the PAR of the solution vectors and their power increase compared to the LS solution, and we have shown that $\ell^\infty$-norm-minimal solutions are optimal under this trade-off. 
We have developed an FBS-based algorithm that is able to efficiently produce minimal-PAR solutions, which are, in general, unattainable by $\ell^\infty$-norm minimization.
In order to demonstrate the efficacy of our approach, we have applied it to joint precoding and PAR reduction (referred to as JPP) in a massive MU-MIMO-OFDM scenario, which has revealed that our new $\ell^p\!-\!\ell^q$-norm formulation is able to outperform CRAMP~\cite{studer15b}, which directly minimizes the $\ell^\infty$-norm.

%% file: 7-appendix.tex
% !TEX root = 21ASIL.tex
% DO NOT REMOVE THE ABOVE COMMENT!

\appendices

\section{Proof of \fref{lem:PARPINCtradeoff}}
\label{app:PARPINCtradeoff}
We have that
\begin{align}
\textit{PAR}(\bmx) \, \textit{PINC}(\bmx)  & = \frac{N\|\bmx\|_\infty^2}{\|\bmx\|_2^2} \frac{\|\bmx\|_2^2}{\|\hat\bmx^\text{LS}\|_2^2} 
  = \frac{N\|\bmx\|_\infty^2}{\|\hat\bmx^\text{LS}\|_2^2},
\end{align}
since $\bmx\neq\bZero$ as $\bmy\neq\bZero$. 
The numerator is minimized by the $\ell^\infty$-norm solution  $\hat\bmx^\infty$ given by~\fref{eq:linftyminimization}, which leads to 
\begin{align}
\textit{PAR}(\bmx) \, \textit{PINC}(\bmx)  &  \geq  \frac{N\|\hat\bmx^\infty\|_\infty^2}{\|\hat\bmx^\text{LS}\|_2^2}  = c.
\end{align}
This observation also implies that  
\begin{align}
\textit{PAR}(\hat\bmx^\infty ) \, \textit{PINC}(\hat\bmx^\infty ) = c,
\end{align}
meaning that the $\ell^\infty$-norm solution is not only trade-off optimal but also determines the lower bound constant $c$. 

%%%
\section{Proof of \fref{lem:parpqinequality}}
\label{app:parpqinequality}

From \fref{eq:normequivalence}, it follows that  $\|\bmx\|^2_p\leq N^{\frac{2}{p}} \|\bmx\|^2_\infty$.
If $2\leq q$, then we also have $\|\bmx\|^2_2 \leq N^{1-\frac{2}{q}} \|\bmx\|^2_q$, which leads to
\begin{align}
 \frac{N^{\frac{2}{q}-\frac{2}{p}}\|\bmx\|^2_p}{\|\bmx\|^2_q}  \leq \frac{N^{\frac{2}{q}}\|\bmx\|^2_\infty}{N^{\frac{2}{q}-1}\|\bmx\|^2_2} = \frac{N\|\bmx\|^2_\infty}{\|\bmx\|^2_2} = \textit{PAR}(\vecx).
\end{align}
If $q=1,p=2$, then the inequality in \fref{lem:parpqinequality} becomes 
\begin{align}
\frac{N\vecnorm{\vecx}_2^2}{\vecnorm{\vecx}_1^2} \leq \frac{N\vecnorm{\vecx}_\infty^2}{\vecnorm{\vecx}_2^2} \Longleftrightarrow \vecnorm{\vecx}_2^2  \leq {\vecnorm{\vecx}_1}\vecnorm{\vecx}_\infty,
\end{align} 
which holds by H\"older's inequality.
%We have that
%\begin{align}
%\frac{\vecnorm{\vecx}_2^2}{\vecnorm{\vecx}_1} = \frac{\sum_{n=1}^N |x_n|^2}{\sum_{n=1}^N |x_n|} \leq \frac{\sum_{n=1}^N \vecnorm{\vecx}_\infty|x_n|}{\sum_{n=1}^N |x_n|} = \vecnorm{\vecx}_\infty,
%\end{align}
%\begin{align}
%\frac{\vecnorm{\vecx}_2^2}{\vecnorm{\vecx}_1} = \frac{\sum_n |x_n|^2}{\vecnorm{\vecx}_1} \leq \frac{\vecnorm{\vecx}_\infty\sum_n |x_n|}{\vecnorm{\vecx}_1} = \vecnorm{\vecx}_\infty,
%\end{align}